\documentstyle[fleqn]{article}
\title{ Non-Abelian  Kubo  Formula and the
Multiple Time-Scale Method}
\author{Zhang Xiaofei\ \ \ \ \ \ \ Li Jiarong\\
\footnotesize{Institute of Particle Physics, Hua-Zhong Normal University,
 Wuhan 430070, China}}

\begin{document}
\maketitle

\begin{quotation}
\begin{abstract}
The non-Abelian  Kubo formula is derived from the
kinetic theory.  That expression is compared with the one
obtained usingthe eiknoal for a Chern-Simons theory.
The multiple time-scale method  is used to study
the non-Abelian  Kubo formula, and the damping rate for
longitudinal color waves is computed.

\end{abstract}
\end{quotation}
\bigskip
\bigskip

\newpage

\vskip1.5cm

\section*{ 1. Introduction}
Substantive progress is made in  the study of non-equilibrium
quark-gluon  plasma in recent years\cite{NQGP}.
The Kubo formula is a fundamental tool for  describing a non-equilibrium
system. Hence it is important to get the Kubo formula of
a non-Abelian  system. In Ref.[2], meaningful results  are given.
In this paper, we derive  a form of the  non-Abelian generation of
the Kubo formula  from  kinetic theory directly,
and the multiple time-scale  method is used to investigate it.
To show the intention of our work concretely, we first give the result
in Ref.[2].  Using the relation between the current and  the generating
fuctional, i.e., $j^\mu(x)=-\delta \Gamma[A]/\delta A_\mu(x)$,
and the Yang-Mills equation,
a form of non-Abelian  Kubo formula is  obtained with
the color current being   expressed
in terms of the eikonal for a chern-Simons theory\cite{HCS}:
\begin{equation}
\label{mf}
\bigl[D_{\nu}F^{\nu\mu}(x)\bigr]_a=j_a^{\mu}(x),
\end{equation}
where $D_{\nu}=\partial _{\nu}-ig[A_{\nu}(x),\cdots]$,
\begin{equation}
\label{ku01}
j^\mu_a(x)={1\over (2\pi)^4}\int d^4k e^{-ik\cdot x}j^{\mu}_a(k)
          ={1\over (2\pi)^4}\int d^4k e^{-ik\cdot x}
          \sum\limits_{n=1}^{\infty} j^{\mu}_{n,a}(k),
\end{equation}
and $j^{\mu}_{n,a}(k)$ is expressed in terms of powers of $A_\mu$,
\begin{eqnarray}
\label{ckua}
j^{\mu}_{n,a}(k)=&-(N_c+{N_f\over 2}){T^2\over 12\pi}g^2
\sum\limits_{k_1+\cdots+ k_n=k}\int d\Omega\biggl[
{\rm tr}\bigl\{  (I_a v^\mu) (A(k_1)\cdot v)  \bigr\}\delta_{n1}\nonumber\\
&-g^{n-1} {\rm tr}\biggl\{ (I_a v^\mu) (A(k_1)\cdot v)\cdots
(A(k_n)\cdot v)  \biggr\} F(k_1,\cdots, k_n)\biggr],
\end{eqnarray}
\begin{eqnarray}
\label{ckub}
F(k_1,\cdots, k_n)=\sum\limits_{i=0}^n{-q_i\over ({\bar q_0 }-{\bar q_i })
 ({\bar q_1 }-{\bar q_i })\cdots  ({\bar q_{i-1} }-{\bar q_i })
({\bar q_{i+1} }-{\bar q_i }) \cdots   ({\bar q_{n} }-{\bar q_i })     },
\end{eqnarray}
where $v=(1,{\bf v})$ is a four light-like vector, and $v'=(1, -\bf v)$,
${\bar q}_0=0$, $ {\bar q}_i=\sum\limits_{j=0}^i (k_j\cdot v+i0^+)$,
$q_i=\sum\limits_{j=0}^i(k_j\cdot v')$.
$N_f$ being the number of flavors for massless quarks,
$I_a $ is the SU($N_c$) group generator in the fundamental representation.
$d \Omega $  denotes integration
over angular direction of the unit vector $\bf v$.
In the above expression,  the relation
$\int d\Omega f[A\cdot v]=\int d\Omega f[A\cdot v']$ has been taken consider of,
and  we have defined,
\begin{equation}
\sum\limits_{k_1+\cdots+ k_n=k}=\int{ d^4k_1\over (2\pi)^4}\cdots
\int{d^4k_n\over (2\pi)^4}(2\pi)^4\delta^{(4)}(k-k_1-\cdots -k_n).
\end{equation}

The non-Abelian  Kubo formula describes Landau damping
of  fields in a quark-gluon plasma as well as Debye screening
and progration of plasma waves in a gauge covariant way.
Finding solutions to the non-Abelian Kubo formula is appealing,
since such solutions  give  collective effects of
the quark-gluon plasma at high temperature. R. Jackiw, Q.Liu and C. Lucchesi
investigate the  static non-Abelian  Kubo formula and show that there are
no hard thermal solitons\cite{stsu}. However, the above expression  of the Kubo
formula is quite complicated, this makes it much more difficult  to use it
to analysis  relativistic physical problems.
Recently, P. F. Kelly, Q.Liu   C. Lucchesi and C. Manuel
 have shown that the generating functional  of hard thermal
loops can be derived from classical transport theory\cite{CH},
and it has been shown  that the  solution of the kinetic equations
taking into account the
non-Abelian and nonlinear effects   can
be obtained using the multiple time-scale method\cite{ZL}.
 The multiple time-scale method is   one
of the  most powerful asymptotic methods in solving  nonlinear   equations,
and it  is proved successful in studying  nonlinear properties of
electromagnetic plasmas [7].
This gives us a chance of investigating the non-Abelian  Kubo
formula from  kinetic theory.
In this paper,    the non-Abelian
Kubo formula  will be derived from the kinetic theory,
then  the multiple time-scale method will be  used to  study  it
and damping rate  will be calculated.

  This paper is organized as follows:
In section 2,  the non-Abelian  Kubo formula is derived from
the kinetic theory and the identity of it with the form obtained using the
eiknoal for a Chern-Simons theory is shown.
In section 3,
the multiple time-scale method  is introduced to study the non-Abelian generation
of the Kubo formula.
In section 4, using the multiple time-scale method, the damping rate
due to non-Abelian and nonlinear nature of QGP eigenwaves
is calculated in the temporal gauge for   longitudinal color waves,
 and the result is compared with the hard thermal
result in thermal QCD.

\section*{2. Non-Abelian  Kubo formula  derived  from the
kinetic approach}

  In this section, we will derive the non-Abelian  Kubo formula
from the kinetic approach.
The phase space of the distribution function of colored particles
is $(x,p, Q)$,
where $Q^a, a=1,2\cdots N^2-1$  are color charges,
and the use of the color charge as
phase-space coordinates is justified recently[5].
Taking cosider of physical constraints on
the phase space volume elemeant $dx dPdQ$, the momentum and
the color charges measure
\begin{equation}
dP={d^4P\over (2\pi)^3}2\theta(p^0)\delta (p^2-m^2),
\end{equation}
\begin{equation}
dQ=d^8Q\delta({Q_aQ^a-q_2})\delta({d_{abc}Q^aQ^bQ^c-q_3}),
\end{equation}
where the constants $q_2$ and $q_3$ fix the values of the Casimir
invariants and $d_{abc}$ are the totally symmetric group constants.

When the classical transport equations for  colored particles is expanded
consistently in the coupling constant g, at the leading order,
the transport equations for the distribution functon $f(x,p,Q)$
is reduced to(neglecting spin effects)[5]
\begin{equation}p^{\mu}\bigl[ \partial_\mu
-gf^{abc}A_\mu^bQ^c\bigr]f(x,p,Q)
=-gQ_aF^a_{\mu\nu}(x){\partial\over \partial p_\nu}f^{(0)}(p),
\end{equation}
where  distribution functions $f^{(0)}(p)$  in the above equation
is the equilibrium distribution function in the absence of a net
color color field,
 \begin{equation}
f^{(0)}(p)=
=C n_{B,F},
\end{equation}
where C is a normalization constant, $n_{B,F}$  are
Fermi-Dirac and Bose-Einstein equilibrium distribution function respectively
\begin{equation}
n_{F}
=(e^{\beta p\cdot U}+1)^{-1},
\ \ \ \ \   \ \ \ \ \ \ \ \ \
n_{B}=(e^{\beta p\cdot U}-1)^{-1},
\end{equation}
where $\beta $  is  the inverse of  the system temperature $T$, $U^\mu$
is the local four  velocity.

The   Yang-Mills mean  field  equation coupled  with the transport equation is
\begin{equation}
\bigl[D_{\nu}F^{\nu\mu}(x)\bigr]_a=j_a^{\mu}(x),
\end{equation}
where  color current $j^{\nu}$  generated  by quarks and gluons
is expressed as,
\begin{eqnarray}
j^{\nu}(x)=\sum_{\rm species}\sum_{\rm helicites}\int dP dQQ^af(x,p,Q)
\end{eqnarray}
Using the above equation, the constraint on
the first order current  demanding by the leading order transport equation
can be obtained\cite{CH}
\begin{equation}
\label{ku1}
[p\cdot D J^\mu(x,p) ]_a=2g^2p^\mu p^\nu F_{\nu 0}^a(x){d\over dp_0}
[N_f f^{(0)}(p)+N_c G^{(0)}(p)],
\end{equation}
where the  field tensor is only related to $F_{\nu 0}$.

The constraint equation (13) can lead  the generating functional of hard
thermal loops\cite{CH}. This means that  Eq.(13) determines
the color current generated by
hard thermal loops and can be used to study the response of quark-gluon
plasma to  fields consistently. In the following discussion, we will use
the constraint equation (13) to get the expression for the color current.

In  the momentum space the  constraint equation (~\ref{ku1})  yields,
\begin{eqnarray}
\label{ku2}
&&(p\cdot k)J^{\mu}_a(k,p)+igf^{abc}\sum\limits_{k'+k''=k}(p\cdot A(k'))_bJ^\mu_c(k'',p)\nonumber\\
&&\ \ =2g^2p^\mu[(p\cdot k)g_0^\nu-\omega p^\nu]A_\nu^a(k){d\over dp^0}[N_c G^{(0)}(p)
+N_f f^{(0)}(p)]\nonumber\\
&&\ \ +2ig^3p^\mu p^\nu \sum\limits_{k_1+k_2=k}f^{abc}A_\nu^b(k_1)A_0^c(k_2)
{d\over dp^0}[N_c G^{(0)}(p)+N_f f^{(0)}(p)].
\end{eqnarray}

 The color current can be expressed as a  series of powers in $A_\mu$,
and in the above equation there is no constraint on the  field potential $A_\mu$,
thus Eq.(~\ref{ku2}) can  be reduced to the  equations of every power in $A_\mu$.
After integrating over the momentum  space one can get
\begin{equation}
\label{ku5}
j^\mu_{1,a}(k)=-m_D^2[g_0^\mu g_0^\nu-\int {d \Omega\over 4\pi}
{\omega v^\mu v^\nu\over v\cdot k +i0^+} ]\   A_\nu^a(k),
\end{equation}
\begin{eqnarray}
\label{ku6}
j^\mu_{2,a}(k)=&&-igm_D^2 f^{abc}\int {d \Omega\over 4\pi}
\sum\limits_{k_1+k_2=k}
{\omega_1\over(v\cdot k +i0^+)(v\cdot k_1+i0^+)}\nonumber\\
&&\ \ \ \  v^\mu v^\lambda v^\nu A_\nu^b(k_2) A_\lambda^c(k_1),
\end{eqnarray}
\begin{equation}
\label{ku7}
j^\mu_{n,a}(k)=-igf^{abc}\int d \Omega {1\over v\cdot k+i0^+}\sum\limits_{k'+k''=k}
(v\cdot A(k'))_bj^{\mu,\Omega}_{n-1,c}(k'')
\ \ \ \ \ \ \ \ \ \  n>2,
\end{equation}
where  $m_D^2={1\over 3}(N_c+{1\over 2}N_f)g^2T^2$,
${j^\mu}_{n-1,a}(k)=\int d \Omega j^{\mu,\Omega}_{n-1,a}(k)$
and to avoid the
poles in the above integrand, the retard condition, i.e., replacing $\omega$
by $\omega+i0^+ $, has been imposed.

The expression for the current to any  power of $A_\mu$ can be obtained easily
using the above equation. Further,  $j^\mu_{n,a} $ can be expressed as
\begin{eqnarray}
\label{ku8}
j^\mu_{n,a}(k)=&&-\delta_{n1}m_D^2[g_0^\mu g_0^\nu-\int {d \Omega\over 4\pi}
{\omega v^\mu v^\nu\over v\cdot k +i0^+} ]\   A_\nu^a(k)
\nonumber\\
&&+(-ig)^{n-1}m_D^2\int {d\Omega\over 4\pi }\sum\limits_{k_1+\cdot+ k_n=k}
f^{aa_nb_{n-1}} f^{b_{n-1}a_{n-1}b_{n-2}}\cdots f^{b_2a_2b_1}\nonumber\\
&&  {\omega_1 v^\mu\over( v\cdot k+i0^+)\bigl(v\cdot(k-k_n)+i0^+\bigr)
 \cdots\bigl( v\cdot( k-k_n-\cdots -k_2)+i0^+ \bigr)}\nonumber\\
&&\ \ \ \ \ \ \  (A(k_n)\cdot v)_{a_n}
\cdots(A(k_2)\cdot v)_{a_2}(A(k_1)\cdot v)_{b_1}.
\end{eqnarray}
Using Eq.(2) and the above equation to express the color current in
the Yang-Mills mean field
equation (11),  the non-Abelian  Kubo formula is derived
from the kinetic theory.
The above expression for the color current is relatively simple, plain,
and  the non-Abelian  Kubo formula expressed by it
is easier for being studied. In the case when the  mean field
is weak,  the non-Abelian  Kubo formula  can be solved by
the multiple time-scale method.

Now  we turn to show that the  expression for the color current
obtained from the kinetic theory is  equivalent  to  the one obtained using
the eiknoal for a Chern-Simons theory.

In the kinetic theory, the expression for
the color current is determined   by Eqs.(~\ref{ku5}-~\ref{ku7}).
One  can  check that $j^\mu_{1,a}(k)$ and $j^\mu_{2,a}(k)$
in  Eq.(~\ref{ku5})
and   Eq.(~\ref{ku6})  are equivalent to those  got from Eqs.(3-~\ref{ckub}).
Thus to show the identity of these two
expressions, one only need to show $j^\mu_{n,a}$ in Eqs.(3-~\ref{ckub}) satisfy
Eq.(~\ref{ku7}). Now we turn to show this.

Using the relation
\begin{equation}
{1\over \bar q_i (\bar q_n-\bar q_i)}= {1\over \bar q_i \bar q_n}
+{1\over \bar q_n (\bar q_n-\bar q_i)},
\end{equation}
And taking note of the delta function in the definition (5), one can  get
from Eq.(4),
\begin{equation}
F(k_1,\cdots, k_n)={1\over v\cdot k+i0^+}[F(k_1,\cdots, k_{n-1})
-F(k_2,\cdots, k_n)].
\end{equation}
Then using the invariance  of $j^\mu_{n,a}$  when $k_1,\cdots ,k_n$ are
exchanged, we obtained  from Eq.(3)
\begin{eqnarray}
j^\mu_{n,a}(k)=&&{1\over v\cdot k+i0^+}(N_c+{N_f\over 2}){T^2\over 12\pi}
\sum\limits_{k_1+\cdots+ k_n=k}\int d\Omega
 {\rm tr}\biggl\{ ([I_b,I_a] v^\mu)\nonumber\\
&&\ \ \ \ \ \ \ \ \ \  (A(k_1)\cdot v)\cdots
(A(k_{n-1})\cdot v)  \biggr\} F(k_1,\cdots, k_{n-1})(A(k_n)\cdot v)^b\nonumber\\
=&&-if^{abc}\int d \Omega{1\over v\cdot k+i0^+}
 \sum\limits_{k'+k''=k}(v\cdot A(k'))_bj^{\mu,\Omega}_{n-1,c}(k'').
\end{eqnarray}
This concludes our proof of the identity  of the expression of the
non-Abelian Kubo formula from kinetic theory and the one  obtained
using the eiknoal for a Chern-Simons theory.

In  the following section, we will investigate the non-Abelian
 Kubo formula  obtained from the kinetic theory, i.e., Eq.(11) coupled with Eq.(2) and
Eq.(~\ref{ku8}), using the multiple time-scale method.

\section*{3. The multiple time-scale method}
The multiple time-scale method is  a kind
of  asymptotic methods of  solving  nonlinear   equations.
In  plasma theory, it is used to analysis the nonlinear
effects of a electromagnetic plasma[7]. Generally speaking, when
 the field strength is not very strong, the mean field
equation may be solved by perturbation methods.  However,
As is known, the nonlinear  wave equation in a plasma  can not
be solved by  naive perturbation analysis, which may   yield
singularity in the solutions  by   virtue  of the resonance in the
relevant equations. This difficulty may be overcome if it is solved
by means of the multiple time-scale perturbation approach[7,8].
In this method, first of all, the independent time variable $t$
is extended   to many time variables, each with different scale,
by introducing the new time variables [7]
\begin{equation}
t^{(n)}\equiv\kappa^nt, \ \ \ \ \ \ \ \ n=0,1,2,\cdots,N ,
\end{equation}
 where $\kappa $ is a dimensionless  parameter introduced to denote the
order of a small quantity.
Then,  every   function   dependent on $t$, such as $Y[t]$,
is  expanded into  an asymptotic series of the form
\begin{equation}Y[t,\kappa]=\sum_{n=0}^N\kappa^nY^{(n)}[t^{(0)}, t^{(1)},\cdots,t^{(N)}]
+O(\kappa^{N+1}). \end{equation}

Every function dependent on $t$  is assumed to be a function of the
multiple time, so the  time derivative should be replaced by
\begin{equation}
{\partial \over \partial t}=\sum_{n=0}^N\kappa^n
{\partial \over \partial t^{(n)}}.
\end{equation}

Usually, one prefers to discuss  the nonlinear wave equations in  momentum
space. The multiple time-scale procedure can also be  used there by
introducing  the multiple scale Fourier transformation
\begin{eqnarray}
Y[t^{(0)}, t^{(1)},\cdots,t^{(N)}]&=&\int_{-\infty}^{\infty} \cdots
\int_{-\infty}^{\infty}
Y(\omega^{(0)}, \omega^{(1)},\cdots ,\omega^{(n)})\nonumber\\
&&\exp \biggl(-i\sum_{n=0}^N\omega^{(n)}t^{(n)} \biggr)\prod\limits_{n=0}^N d\omega^{(n)}.
\end{eqnarray}
 Using this transformation,   the multiple time-scale
 method   can be used in  momentum space simply   by taking   the following
 expansions instead of Eqs.(22-24)[8],
\begin{equation}
\label {EX1}
 \omega=\sum_{n=0}^N\kappa^n\omega^{(n)} ,
 \end{equation}
 \begin{equation}
 \label{EX2}
Y[\omega,\kappa]=\sum_{n=0}^N\kappa^nY^{(n)}[\omega^{(0)}, \omega^{(1)},
\cdots,\omega^{(N)}]+O(\kappa^{N+1}).
\end{equation}

In this paper,  we will use the multiple time-scale method to study
the non-Abelian  Kubo formula in  momentum space.
The corresponding equation of  Eq.(11) in  momentum space is obtained
by using
\begin{equation}{ A^\mu}(x)=\int{d^4k\over(2\pi)^4}e^{-ik\cdot x}
{ A^\mu}(k). \end{equation}
The  non-Abelian  Kubo formula in the momentum space is
\begin{eqnarray}
\label{CE1}
&-&(k^2g^{\nu\mu}-k^\nu k^\mu)A_\nu(k)-g\sum\limits_{k_1+k_2=k}(k^\nu+k_2^\nu)
[A_\nu(k_1),A^\mu(k_2)]\nonumber\\
&+&g\sum\limits_{k_1+k_2=k}(k_2^\mu)[A^\nu(k_1),A_\nu(k_2)]
+g^2\sum\limits_{k_1+k_2=k}\sum\limits_{k_3+k_4=k_2}
[A^\nu(k_1),[A_\nu(k_3),A^\mu(k_4)]]\nonumber\\
&=&j^\mu(k),
\end{eqnarray}
where $j^\mu(k) $ is expressed as $Eq.(18)$.

As we have discussed  in the beginning of this section,
when  the mean field is not very strong,
the non-Abelian  Kubo formula
can be solved by the multiple time-scale method.
If we work in the momentum space,
following Eq.(~\ref{EX1}) and Eq.(~\ref{EX2}),
the frequency and the color field potential are expanded into
asymptotic series of forms as
\begin{equation}
\label{EX3}
\omega=\omega^{(0)}+\kappa \omega^{(1)}+\kappa^2 \omega^{(2)}+
\kappa^3\omega^{(3)}+\cdots,
\end{equation}
\begin{eqnarray}
\label{EX4}
A_\mu =&\kappa& A^{(1)}_\mu(\omega^{(0)},\omega^{(1)},\omega^{(2)},
\omega^{(3)},\cdots)
+\kappa^2 A^{(2)}_\mu(\omega^{(0)},\omega^{(1)},\omega^{(2)},
\omega^{(3)},\cdots)\nonumber\\
&+&\kappa^3 A^{(3)}_\mu(\omega^{(0)},\omega^{(1)},\omega^{(2)},
\omega^{(3)},\cdots)+\cdots.
\end{eqnarray}

Inserting the expansions of all the variables into the nonlinear equation,
and   equating the coefficients of corresponding powers of $\kappa$ in
the  two sides of
this equation,  the perturbation expansions of the nonlinear equation
at every order are given. The nonlinear equation then can be discussed
iteratively.

Now, we  first discuss the physical meaning of the above expansions given
by the multiple time-scale method.
In the linear response approximation, any perturbation in a plasma
may be taken to be a superposition of independent eigenwaves
whose dispersion relation is determined by the solution to
the non-Abelian  Kubo  formula in the linear  approximation.
When the nonlinear
terms in the non-Abelian  Kubo formula are retained,
the nonlinear interactions of the eigenwaves will be taken into account.
The nonlinear interactions of the eigenwaves  will change the dispersion
relation  obtained in the linear approximation, i.e., frequency corrections
$\omega^{(1)}, \omega^{(2)} \cdots$  expressed by Eq.(~\ref{EX3}) will be given to
the eigenfrequency $\omega^{(0)}$ correspondingly. If the
imaginary parts of  the frequency corrections for  eigenmodes
are  not vanishing,  there will be instablity or damping for the eigenmodes.
In next section, we will calculate the damping rate due to
interactions of QGP eigenwaves  in the temporal gauge.

\section*{ 4. Nonlinear damping rate  in the temporal gauge}
\setcounter{section}{3}

\ \ \ \ The non-Abelian  Kubo formula determines the dispersion
relation of the  eigenwaves in QGP. The damping rate of  plasma eigenwaves
is determined by the imaginary part of the dispersion relation.  In  this
section we will calculate the damping rate   for longitudinal color eigenwaves
in the long-wavelength limit   from the non-Abelian  Kubo
formula.

In this section, for convenience  we will work  in the temporal gauge,
$A^0_a=0$. The relation between the color electric field ${\bf E}$ and
the vector potential in that gauge is simple,
\begin{equation}
{E^i}_a(x)=-\partial{ A^i}_a(x)/ \partial t.
\end{equation}

 The non-Abelian  Kubo  formula  in the momentum
space and in the  temporal gauge is
\begin{eqnarray}
&&-(\omega^2-K^2)A^h(k)-k^ik^h A^i(k)+g\sum\limits_{k_1+k_2=k}^{ }
k^i[A^i(k_1),A^h(k_2)]\nonumber\\
&&+g\sum\limits_{k_1+k_2=k}^{ } k_{2}^i[A^i(k_1) ,A^h(k_2)]
-g\sum\limits_{k_1+k_2=k}^{ } k_{2}^h[A^i(k_1),A^i(k_2)]\nonumber\\
&&+g^2\sum\limits_{k_1+k_2=k}\sum\limits_{k_3+k_4=k_2}^{ }
[A^i(k_1),[A^i(k_3),A^h(k_4)] ] =j^h(k),
\end{eqnarray}
where $K=|{\bf k}|$ and $j^h(k)$ is expressed as Eq.(18).

Though the method used in this paper is general,
for simplicity of  calculation, we suppose that there is only
longitudinal vector potential, i.e., the vector potential is parallel
to the wave vector,

\begin{equation}
A^i(k)=k^iA(k)/K.
\end{equation}

Inserting  the expansions (~\ref{EX3}-~\ref{EX4}) into Eq.(33) and Eq.(18),
for longitudinal  field,  the first  order  non-Abelian 
Kubo formula is,
\begin{eqnarray}
-({\omega^{(0)}}^2-K^2)A_a^{(1)h}(k)-k^ik^h A_a^{(1)i}(k)
= -m_D^2\int {d \Omega\over 4\pi}
{\omega^{(0)} v^h v^j\over v\cdot k +i0^+}    A^{j(1)}_a(k).
\end{eqnarray}

The above  non-Abelian  Kubo formula can be reduced to
\begin{equation}
\epsilon(\omega^{(0)}, {\bf k})A^{(1)}(k)=0,
\end{equation}
 where $\epsilon$ is the color electric permeability,
\begin{equation}
\label{DIS1}
\epsilon(\omega^{(0)}, {\bf k})=1+{3\omega_p^2\over K^2} [1-{\omega^{(0)}
\over 2K}\bigl({\ln}|{K+\omega^{(0)}\over K-\omega^{(0)}}|
-i\pi\Theta(K-\omega^{(0)}) \bigr) ].
\end{equation}
where $\omega_p^2=(2N_c+N_f)g^2T^2/18$.

 From the above equation, the  solution to the first order non-Abelian
  Kubo formula  can be got
\begin{equation}
\label{A1}
A_{{\bf k}}^{(1)\sigma}=-i{\pi\over \omega^{(0)}}E^{\sigma}_{{\bf k}0}
[e^{-i\phi^\sigma_{\bf k}}
\delta(\omega^{(0)}-\omega_{\bf k}^\sigma)+e^{i\phi^\sigma_{\bf k}}
\delta(\omega^{(0)}+\omega_{\bf k}^\sigma)],
\end{equation}
 where  $E^{\sigma}_{{\bf k}0}$ and $\phi^\sigma_{\bf k}$
are  the initial amplitude and phase of the oscillation respectively
($\sigma $ is introduced to label different eigenwaves).
The frequencies and the wave vectors must satisfy the dispersion relation,
\begin{equation}
\label{DIS2}
\epsilon(\omega^{(0)}, {\bf k})=0.
\end{equation}
Eq.(~\ref{DIS2} ), which is quite similar to
the dispersion relation for the
abelian-like  plasma eigenwaves, is the first order dispersion relation
for the color eigenwaves in a QGP.
It agrees with the leading order result using
the finite temperature QCD\cite{weldon}.
It  has been shown by U.Heinz that  as the  eigenwaves
in QGP are  always time-like, ${\omega^{(0)}}^2>K^2$, the color eigenwaves in
a QGP do not undergo  Landau damping in the linear approximation[10].

  Using the results of the first order equation, the second order equation
now can be considered. The second order  non-Abelian  Kubo
formula for the longitudinal field is
\begin{eqnarray}
\label{3W}
&&-{\omega^{(0)}}^2 \epsilon(\omega^{(0)},{\bf k})A_a^{(2)}(k){k^h \over K}
-2\omega^{(1)}\omega^{(0)}A_a^{(2)}(k){k^h \over K}\nonumber\\
&&-2g\sum\limits_{k_1+k_2=k}{{\bf k}\cdot{\bf k_1} \over K_1} {k_2^h \over K_2}
{\rm tr}\bigl\{I_a[A^{(1)}(k_1),A^{(1)}(k_2)]\bigr\}\nonumber\\
&&=-m_D^2\int {d\Omega\over 4\pi   }{\omega^{(0)}\over K  } v^h A^{(1)}_a(k)
{{\bf v}\cdot {\bf k}\over v\cdot k+i0^+}\nonumber\\
&&-2gm_D^2g\sum\limits_{k_1+k_2=k}
\int {d\Omega\over 4\pi   }
{ {\bf v}\cdot{\bf k}_1 {\bf v}\cdot{\bf k}_2\over K_1 K_2}
{v^h \omega^{(0)}_1 \over(v\cdot k+i0^+) (v\cdot k_1+i0^+)}\nonumber\\
&&{\rm tr}\bigl\{I_a[A^{(1)}(k_2),A^{(1)}(k_1)]\bigr\},
\end{eqnarray}
where the right side of the above equation comes from
the second order
contribution   of $j^\mu_{a,1}$ and $j^\mu_{a,2}$ in Eqs.(15-16).

   Eq.(~\ref{3W}) discribes the nonlinear effects owing  to the three-wave
processes  which are the merging of two eigenwaves with eigenfrequencies
$ \omega^{(0)\sigma}_{{\bf k}_1}$ , $\omega^{(0)\lambda}_{{\bf k}_2}$  into a
third wave, or  the corresponding inverse processes.
The appearance of the term $[A^{(1)}(k_1),A^{(1)}(k_2)]$
in Eq.(~\ref{3W}) indicates that the non-Abelian nature of the mean color field
has been taken into account and the interaction of the QGP eigenwaves.

 Now, we  consider the nonlinear effects due to the three-wave interactions.
We will see that the nonlinear effects owing to the three-wave interactions
on the eigenfrequency are vanishing. In the three-wave processes, when the
third wave is also an eigenwave, this kind of processes is called three-wave
resonance. Taking note of Eq.(5), the following matching condition
representing the conservation of energy should be satisfied for the
three-wave resonance,
\begin{equation}
\label{MC}
| \omega^{(0)\sigma}_{{\bf k}_1} \pm \omega^{(0)\lambda}_{{\bf k}_2}|
 =\omega^{(0)\tau}_{{\bf k}},
 \end{equation}
  where $\sigma$, $\lambda$, and $\tau$ in the above equation indicate that the
 corresponding frequency is  the eigenfrequency of a certain eigenmode.

The three-wave resonance processes require that  the matching  condition
Eq.(~\ref{MC})  must be satisfied  strictly; even a small mismatch of
Eq.(~\ref{MC})
may significantly decrease the efficiency of the three-wave resonance[11].
This condition limits the permissible region of the three-wave resonance
interaction in $\bf k$ space. Indeed for some forms of  dispersion relation,
the three-wave  resonance is completely forbidden.
It can be proved that
for  dispersion Eq.(~\ref{DIS1}) and Eq.(~\ref{DIS2}), where  the phase velocity decreases with  $K$
[1,10],  the three-wave resonance is forbidden \cite{TP}.
The absence of the three-wave
resonance here leads  the eigenfrequency correction $\omega^{(1)}$ to
vanish, which is similar to the corresponding situation in an
electromagnetic plasma[8]. Using this result and taking note of  Eq.(38),
the second order field potential
is  expressed as
\begin{eqnarray}
\label{EE2}
&&{\omega^{(0)}}^2\epsilon(\omega^{(0)},{\bf k})A^{(2)}(k)\nonumber\\
&&= -2gm_D^2g\sum\limits_{k_1+k_2=k}
\int {d\Omega\over 4\pi   }
{ {\bf v}\cdot{\bf k}_1 {\bf v}\cdot{\bf k}_2\over K_1 K_2}
{v^h \omega^{(0)}_1 \over(v\cdot k+i0^+) (v\cdot k_1+i0^+)}\nonumber\\
&&{\rm tr}\bigl\{I_a[A^{(1)}(k_2),A^{(1)}(k_1)]\bigr\}.
\end{eqnarray}

 From the above discussion,  we can see  that there is no  effect
of three-wave resonance  on the eigenfrequencies. Hence the next order
perturbation equation should be considered  in order to obtain non-vanishing
effects of the non-Abelian and nonlinear wave interactions on the
eigenfrequencies.

Using the above result Eq.(~\ref{EE2}) and $\omega^{(1)}=0$,
the third order  non-Abelian  Kubo formula for
the longitudinal field  is expressed as
\begin{eqnarray}
\label{3EQ}
&&-{\omega^{(0)}}^2\epsilon(\omega^{(0)},{\bf k}) A_a^{(3)}(k)-2\omega^{(2)}\omega^{(0)}A_a^{(1)}(k)
\nonumber\\
&&+2g^2\sum\limits_{k_1+k_2=k}\sum\limits_{k_3+k_4=k_2}
{{\bf k}_1\cdot{\bf k}_3 \over K_1 K_3} {{\bf k}_4 \cdot{\bf k}
\over K K_4}
{\rm tr}\bigl\{I_a[A^{(1)}(k_1),[A^{(1)}(k_3),A^{(1)}(k_4)]]\bigr\}\nonumber\\
&&=-2g^2m_D^4\sum\limits_{k_1+k_2=k}\sum\limits_{k_3+k_4=k_2}g^6
\int d \Omega{1\over v\cdot k^{(0)} +i0^+}
{{\bf v} \cdot {\bf k}\over K}{{\bf v} \cdot {\bf k}_1\over K_1}
{{\bf v} \cdot {\bf k}_2\over K_2}\nonumber\\
&&({\omega^{(0)}_2\over v\cdot k_2^{(0)} +i 0^+}
-{\omega_1^{(0)}\over v\cdot k_1^{(0)} +i 0^+})
{1\over{\omega^{(0)}_2}^2}{1\over\epsilon(k_2^{(0)})}\int d \Omega'
{1\over v'\cdot k_2^{(0)} +i0^+}
\nonumber\\
&&{\omega_4^{(0)}\over v'\cdot k_4^{(0)} +i0^+}{{\bf v}' \cdot {\bf k}_2\over K_2}
{{\bf v}' \cdot {\bf k}_3\over K_3}
{{\bf v}' \cdot {\bf k}_4\over K_4}
{\rm tr}\bigl\{I_a[A^{(1)}(k_1),[A^{(1)}(k_3) ,A^{(1)}(k_4) ]]\bigr\}\nonumber\\
&&-2g^2m_D^2\sum\limits_{k_1+k_2=k}\sum\limits_{k_3+k_4=k_2} g^4 \int d\Omega
{1\over v\cdot k^{(0)} +i0^+}{{\bf v} \cdot {\bf k}\over K}
{{\bf v} \cdot {\bf k}_3\over K_3}{{\bf v} \cdot {\bf k}_1\over K_1}
{{\bf v} \cdot {\bf k}_4\over K_4}\nonumber\\
&&{1\over v\cdot k_2^{(0)} +i0^+}
{\omega_4^{(0)}\over v\cdot k_4^{(0)} +i 0^+}
{\rm tr}\bigl\{I_a[A^{(1)}(k_1),[ A^{(1)}(k_3),A^{(1)}(k_4)] ]\bigr\}.
\end{eqnarray}

  We only  need to discuss the four-wave resonance processes,  as we are
interested in the effects of wave interactions on the
eigenfrequencies.  Therefore,  $k^{(0)}$  in Eq.(~\ref{3EQ} ) satisfies
the dispersion relation Eq.(~\ref{DIS1}) and Eq.(~\ref{DIS2} )
in the following discussion.
In this section, we only calculate damping rate being the imaginary part
of the eigenfrequency, i.e.,
$\gamma=-{\rm Im \omega^{(2)}}$.
Since only
the imaginary  part of Eq.(~\ref{3EQ} ) is related to  the damping rate, it is also
sufficient to consider the imaginary part of Eq.(~\ref{3EQ} )

   As the oscillations are developed from random thermal motions, the
eigenmodes can be considered completely incoherant to each other.
For a  color field being in  random phase, we have
\begin{equation}
\label{RP1}
\langle A^{(1)a}(k)\rangle=0.
\end{equation}
 where $\langle  \ \rangle $ means the  average with respect to the random
phase of the oscillations. Then, using Eq.(~\ref{A1}), one can obtain
\begin{equation}
\label{RP2a}
\langle A^{(1)a}(k){A^{(1)b}}^*(k')\rangle =(2\pi)^4\delta^{(4)}(k-k')\delta_a^b
\langle {A^{(1)}}^2\rangle_{{\bf k}\omega} ,
\end{equation}
\begin{equation}
\label{RP2b}
\langle {A^{(1)}}^2\rangle_{{\bf k}\omega}  ={\pi\over{\omega^{(0)}}^2}
[\delta(\omega^{(0)}-\omega_{\bf k})
+\delta(\omega^{(0)}+\omega_{\bf k})]I_{\bf k},
\end{equation}
\begin{equation}
\label{RP2c}
I_{\bf k}={|E_{\bf k}|^2\over 2V},
\end{equation}
 where $I_{\bf k}$ characterizes the total intensity of the fluctuation
oscillation with frequency $\omega_{\bf k}$ and $-\omega_{\bf k}$,
$V$ is the volume of the plasma.

As the oscillations are incoherent to each other,
the average of product of the field potential can be expanded as
\begin{eqnarray}
\label{RP3}
&&\langle A^{(1)}(k_1)A^{(1)}(k_2)A^{(1)}(k_3)A^{(1)}(k_4) \rangle\nonumber\\
&&=\langle A^{(1)}(k_1)A^{(1)}(k_2)\rangle  \langle A^{(1)}(k_3)A^{(1)}(k_4)\rangle
+\langle A^{(1)}(k_1)A^{(1)}(k_3)\rangle \langle A^{(1)}(k_2)A^{(1)}(k_4)\rangle  \nonumber\\
&&+\langle A^{(1)}(k_1)A^{(1)}(k_4)\rangle \langle A^{(1)}(k_2)A^{(1)}(k_3)\rangle.
\end{eqnarray}

  Multiplying  both sides of Eq.(~\ref{3EQ}) by $A^{(1)b}(k')$, then performing
the average of the result with respect to the random phase using Eq.(~\ref{RP1})
and Eq.(~\ref{RP3}), one can obtain the damping rate
of the longitudinal eigenwaves in QGP
\begin{eqnarray}
\label{DP3}
&&\gamma (K)
=-{\rm Im }\omega^{(2)} \nonumber\\
\ \ \ &&={3m_D^2g^2\over \omega_p}{\rm Im}\biggl\{\int {d\Omega\over 4\pi}
\int {d^4k_1\over (2\pi)^4}\langle {A^{(1)}}^2\rangle_{{\bf k}_1\omega}
{1\over v\cdot k^{(0)}+i0^+}\nonumber\\
\ \ \ && \bigl[
({{\bf v}\cdot{\bf k}\over K})^2
({{\bf v}\cdot{\bf k}_1\over K_1})^2
{1\over v\cdot (k^{(0)}-k_1^{(0)})+i0^+}
({\omega^{(0)}_1\over v\cdot k^{(0)}_1-i0^+}
-{\omega^{(0)}\over v\cdot k^{(0)}+i0^+})\nonumber\\
&&\ \ \ \ +m_D^2({{\bf v}\cdot{\bf k}\over K})
({{\bf v}\cdot{\bf k}_1\over K_1})({{\bf v}\cdot({\bf k}-{\bf k}_1)\over |{\bf k}-{\bf k}_1|})
({\omega^{(0)}-\omega^{(0)}_1\over v\cdot (k^{(0)}-k_1^{(0)})+i0^+}
-{\omega^{(0)}_1\over v\cdot k_1^{(0)}+i0^+})\nonumber\\
&&\ \ \ \int {d\Omega'\over 4\pi}
({{\bf v'}\cdot({\bf k}-{\bf k}_1)\over |{\bf k}-{\bf k}_1|})
({{\bf v'}\cdot{\bf k}\over K})({{\bf v'}\cdot{\bf k}_1\over K_1})
{1\over \epsilon(\omega^{(0)}-\omega^{(0)}_1,
{\bf k}-{\bf k}_1 )}{1\over(\omega^{(0)}-\omega^{(0)}_1)^2}\nonumber\\
&&\ \ \ \
{1\over v'\cdot (k^{(0)}-k_1^{(0)})+i0^+}
({\omega^{(0)}_1\over v'\cdot k_1^{(0)}-i0^+}-
{\omega^{(0)}\over v'\cdot k^{(0)}+i0^+})\bigr]\biggr\}.\nonumber\\
\end{eqnarray}

    Now we perform the integrals in Eq.(~\ref{DP3} ) in cylindrical coordinates
and in plasma particle local rest frame.  For simplicity of calculation the
direction of
polar axis is selected  to be  the same as the direction of ${\bf k}_1$; then
the coordinates for  ${\bf k}$, ${\bf p}$, ${\bf p}'$ are
${\bf k}(K,\alpha,\beta)$, ${\bf p}(E_p,\theta,\varphi)$,
${\bf p}'(E_p',\theta',\varphi')$ correspondingly.
 As   only the damping rate in the long-wavelength
limit is calculated  numerically later,
we will use the approximate relation in the long-wavelength region
in the following calculation.
In the long-wavelength region, the dispersion relation
Eq.(39)  is reduced to
the form of[1,10]
\begin{equation}
\label{DL}
{\omega^{(0)}}^2=\omega_p^2+{3\over 5}K^2.
\end{equation}

For the long-wavelength limit case($K=0$),
using the equation
$p^0/( p\cdot k' +i p^00^+)
=P[1/(\omega'-{\bf v}\cdot {\bf k'})]
-i\pi\delta(\omega'-{\bf v}\cdot {\bf k'})$,
where ${\bf v}={\bf p}/ p^0$ and $P$ stands for principal value
of the function, one can check that the condition for nonvanishing of
Eq.(49) is the satisfation of
the delta function $\delta[\omega^{(0)}-\omega_1^{(0)}
-{\bf v}\cdot({\bf k}-{\bf k}_1)] $.
This delta function  restricts
the upper bound of the integrals over $K_1$  in Eq.(49) to
equal  $gT$ approximately.
We obtain from Eq.(~\ref{DP3})  the expression  of the
damping rate  for ${\bf k}=0$ modes,
\begin{eqnarray}
\label{DP4}
&&\gamma={3g^2m_D^2\over 32\omega_p^2 \pi^2}\int d\omega^{(0)}_1\int_0^{gT}dK_1
I_{{\bf k}_1}
{K_1\over{ \omega^{(0)}_1}^2}({\omega^{(0)}_1\over \omega_p}-1)
\delta [\omega^{(0)}_1-\sqrt{\omega_p^2+{3\over 5}K_1^2}]\nonumber\\
&&\int_{-1}^1d(cos\theta)\int_{0}^{2\pi} d\varphi
\delta[cos\theta-{(\omega_1^{(0)}-\omega_p)\over K_1}]
\int {d \Omega \over 4\pi }\nonumber\\
&&\ \ \ \biggl\{cos^2\phi cos^2\alpha
+{m_D^2\over 4}\int_{-1}^1 d(cos\theta')\int_{0}^{2\pi}
d\varphi'{ cos^2\theta cos^2\theta' \over(\omega_p-\omega^{(0)}_1)^2
\epsilon(\omega_p-\omega^{(0)}_1,-{\bf k}_1 )}
\nonumber\\
&&\ \ \ \ cos\phi cos\phi'\bigl[\bigl({\omega_1^{(0)}\over\omega_p}-1+
{K_1 cos\theta'\over\omega^{(0)}_1-K_1 cos\theta'}\bigr)
{\omega^{(0)}_1-\omega_p\over\omega_p-\omega^{(0)}_1+K_1cos\theta'}\nonumber\\
&&\ \ \ \ \ +{\omega^{(0)}_1\over\omega^{(0)}_1-K_1cos\theta'}({\omega_1^{(0)}\over\omega_p}-1)
\bigr] \biggr\},
\end{eqnarray}
 where $\phi$, $\phi'$, are the  corresponding angles between
$\bf p$ and ${\bf k}$ ,
$\bf p'$ and $\bf k$. They can be expressed as
\begin{equation}
\cos\phi=sin\theta sin\alpha cos(\varphi-\beta)+cos\theta cos\alpha.
\end{equation}
  And the expression for $cos\phi'$ is similar to the above equation.

In Eqs.( ~\ref{DP4}), the  damping rate is
proportional to the wave intensity $I_{\bf k}$. In  an equilibrium plasma,
the wave intensity is governed by the plasma temperture[7][13],
\begin{equation}
I_{\bf k}=8\pi{T\over \epsilon'_{\bf k}\omega _{\bf k}},
\end{equation}
\begin{equation}
\epsilon'_{\bf k}={\partial {\rm Re}\epsilon(\omega,{\bf k})\over \partial
\omega _{\bf k}}.
\end{equation}
For the long-wavelength case, using Eq.(50), we  can obtain
\begin{equation}
I_{\bf k}=4\pi T.
\end{equation}

Then,
after finishing  all the integrals in Eq.(~\ref{DP4}),
the numerical result of the  damping  rate
for pure gluon gas  is obtained
\begin{equation}
\gamma\sim 0.21 g^2T.
\end{equation}

Now, compare the above result with  the damping rate for
long-wavelength gluons calculated by  the effective perturbation theory based
on the resummation  of the hard thermal loops. The hard thermal result is [14]
\begin{equation}
\gamma\sim 0.26 g^2T.
\end{equation}
 Note that both of them are obtained approximately,
we can conclude that they are coincide.
This indicates the consistency between the high temperature
limit of  thermal QCD  and the non-Abelian  Kubo formula
derived from the  kinetic theory for quarks and gluons.

 In this paper, we start from the non-Abelian Kubo formula
  which  is gauge invariant,  and we chose a physical gauge
 to express the gauge invarant equation,
 then the multiple time-scale method is used to expand the  equation
 according the field strength.  Our  result is coincide with the
 gauge invariant result obtained using the effective perturbation theory.
 As to the question if the multiple method guarantees the gauge symmtry
 generally, this need further study.

In conclusion: in this paper we derive the non-Abelian generation of
the Kubo formula from the kinetic theory and show it is coincide with
the form obtained using the eiknoal for a Chern-Simons theory.
The multiple time-scale method is used to solve the non-Abelian Kubo
formula obtained in this paper to calculate the  damping rate for the
eigenwaves in the long-wavelength limit numerically,

the result is coincide with that obtained using the effective perturbation
theory in the finite temperature field theory. This show that the multiple
time-scale method is effective on solving the non-Abelian generation of
the Kubo formula.

\section*{Acknowledgement}

This work is supported in part by the National Nature Science Fund
 of China.

\end{document}